\documentclass[12pt]{article}
\newcommand{\ba}{\begin{eqnarray}}
\newcommand{\ea}{\end{eqnarray}}
\usepackage{epsf,rotate}

\catcode`\@=11
\long\def\@makefntext#1{
\protect\noindent \hbox to 3.2pt {\hskip-.9pt
$^{{\ninerm\@thefnmark}}$\hfil}#1\hfill}                

\def\@makefnmark{\hbox to 0pt{$^{\@thefnmark}$\hss}}  

\def\ps@myheadings{\let\@mkboth\@gobbletwo
\def\@oddhead{\hbox{}
\rightmark\hfil\ninerm\thepage}
\def\@oddfoot{}\def\@evenhead{\ninerm\thepage\hfil
\leftmark\hbox{}}\def\@evenfoot{}
\def\sectionmark##1{}\def\subsectionmark##1{}}

\renewcommand{\thefootnote}{\fnsymbol{footnote}}

\def\sectionc{\@startsection {section}{1}{\z@}{-3.5ex plus -1ex minus
    -.2ex}{2.3ex plus .2ex}{\bf }}
\def\subsectionc{\@startsection{subsection}{2}{\z@}{-3.25ex plus -1ex minus
   -.2ex}{1.5ex plus .2ex}{\it }}
\renewcommand{\section}[1]{\sectionc{#1}\hspace*{\parindent}}
\renewcommand{\subsection}[1]{\subsectionc{#1}\hspace*{\parindent}}
\newcounter{appendixc}
\newcounter{subappendixc}[appendixc]
\newcounter{subsubappendixc}[subappendixc]

\renewcommand{\appendix}[1] {\vspace*{0.6cm}
        \refstepcounter{appendixc}
        \setcounter{figure}{0}
        \setcounter{table}{0}
        \setcounter{equation}{0}
        \renewcommand{\thefigure}{\Alph{appendixc}.\arabic{figure}}
        \renewcommand{\thetable}{\Alph{appendixc}.\arabic{table}}
        \renewcommand{\theappendixc}{\Alph{appendixc}}
        \renewcommand{\theequation}{\Alph{appendixc}.\arabic{equation}}
        \noindent{\bf Appendix \theappendixc #1}\par\vspace*{0.4cm}}

\def\abstracts#1{{

\centering{\begin{minipage}{13.2truecm}\footnotesize
\baselineskip=13pt\noindent
        \parindent=0pt #1
        \end{minipage}}\par}}

\renewenvironment{thebibliography}[1]
        {\begin{list}{\arabic{enumi}.}
        {\usecounter{enumi}\setlength{\parsep}{0pt}
\setlength{\leftmargin 0.75cm}{\rightmargin 0pt}
         \setlength{\itemsep}{0pt} \settowidth
        {\labelwidth}{#1.}\sloppy}}{\end{list}}

\newcounter{itemlistc}
\newcounter{romanlistc}
\newcounter{alphlistc}
\newcounter{arabiclistc}

\newcommand{\fcaption}[1]{
        \refstepcounter{figure}
        \setbox\@tempboxa = \hbox{\footnotesize Figure~\thefigure. #1}
        \ifdim \wd\@tempboxa > 6in
           {\begin{center}
        \parbox{6in}{\footnotesize\baselineskip=13pt Figure~\thefigure. #1}
            \end{center}}
        \else
             {\begin{center}
             {\footnotesize Figure~\thefigure. #1}
              \end{center}}
        \fi}

\newcommand{\tcaption}[1]{
        \refstepcounter{table}
        \setbox\@tempboxa = \hbox{\footnotesize Table~\thetable. #1}
        \ifdim \wd\@tempboxa > 6in
           {\begin{center}
        \parbox{6in}{\footnotesize\baselineskip=13pt Table~\thetable. #1}
            \end{center}}
        \else
             {\begin{center}
             {\footnotesize Table~\thetable. #1}
              \end{center}}
        \fi}

\def\@citex[#1]#2{\if@filesw\immediate\write\@auxout
        {\string\citation{#2}}\fi
\def\@citea{}\@cite{\@for\@citeb:=#2\do
        {\@citea\def\@citea{,}\@ifundefined
        {b@\@citeb}{{\bf ?}\@warning
        {Citation `\@citeb' on page \thepage \space undefined}}
        {\csname b@\@citeb\endcsname}}}{#1}}

\newif\if@cghi
\def\cite{\@cghitrue\@ifnextchar [{\@tempswatrue
        \@citex}{\@tempswafalse\@citex[]}}
\def\citelow{\@cghifalse\@ifnextchar [{\@tempswatrue
        \@citex}{\@tempswafalse\@citex[]}}
\def\@cite#1#2{{$\null^{#1}$\if@tempswa\typeout
        {IJCGA warning: optional citation argument
        ignored: `#2'} \fi}}

 1
 1
 1

\font\ninerm=cmr9

\begin{document}

\centerline{\normalsize\bf RECENT CALCULATIONS OF ELECTROMAGNETIC}
\baselineskip=15pt
\centerline{\normalsize\bf AND STRONG DECAYS OF N*}

\vspace*{0.6cm}
\centerline{\footnotesize A. LEVIATAN$^a$ and R. BIJKER$^b$}
\baselineskip=13pt
\centerline{\footnotesize\it $^a$Racah Institute of Physics, 
The Hebrew University, Jerusalem 91904, Israel}
\centerline{\footnotesize\it $^b$Instituto de Ciencias Nucleares, 
U.N.A.M., A.P. 70-543, 04510 M\'exico, D.F., M\'exico}

\vspace*{0.6cm}
\abstracts{
We report on recent calculations of electromagnetic elastic form factors,
helicity amplitudes and strong decay widths of N* resonances. The 
calculations are done in a collective constituent model for the nucleon, 
in which the resonances are interpreted as rotations and vibrations
of an oblate top with a prescribed distribution of charges and 
magnetization.}

\normalsize\baselineskip=15pt
\setcounter{footnote}{0}
\renewcommand{\thefootnote}{\alph{footnote}}

\section{Introduction}\label{sec:introduction}
Effective models of baryons based on three constituents
share a common spin-flavor-color structure but differ in their 
assumptions on the spatial dynamics. 
Quark potential models in nonrelativistic\cite{NRQM} or relativized
\cite{RQM} forms emphasize the single-particle aspects of quark dynamics 
for which only a few low-lying configurations in the confining potential 
contribute significantly to the eigenstates of the Hamiltonian. 
On the other hand,
some regularities
in the observed spectra ({\it e.g.} linear Regge trajectories,
parity doubling) hint that an alternative, collective
type of dynamics may play a role in the structure of baryons. In this 
contribution we present a particular collective model of baryons\cite{BIL}
and report on calculations of electromagnetic\cite{emff} and
strong\cite{strong} couplings within this framework.

\section{A Collective Model of Baryons}\label{sec:collmodel}
We consider a collective model in which the baryon resonances
are interpreted in terms of rotations and vibrations of the string 
configuration in Fig. 1. The corresponding oblate top wave functions are 
spread over many oscillator shells and hence 
are truly collective. 
A typical mass spectrum in the collective model
is shown in Fig.~2\\
\noindent\begin{tabular}{p{15cm}}
\epsfxsize 1.75in
\rotate{\rotate{\rotate{\epsffile{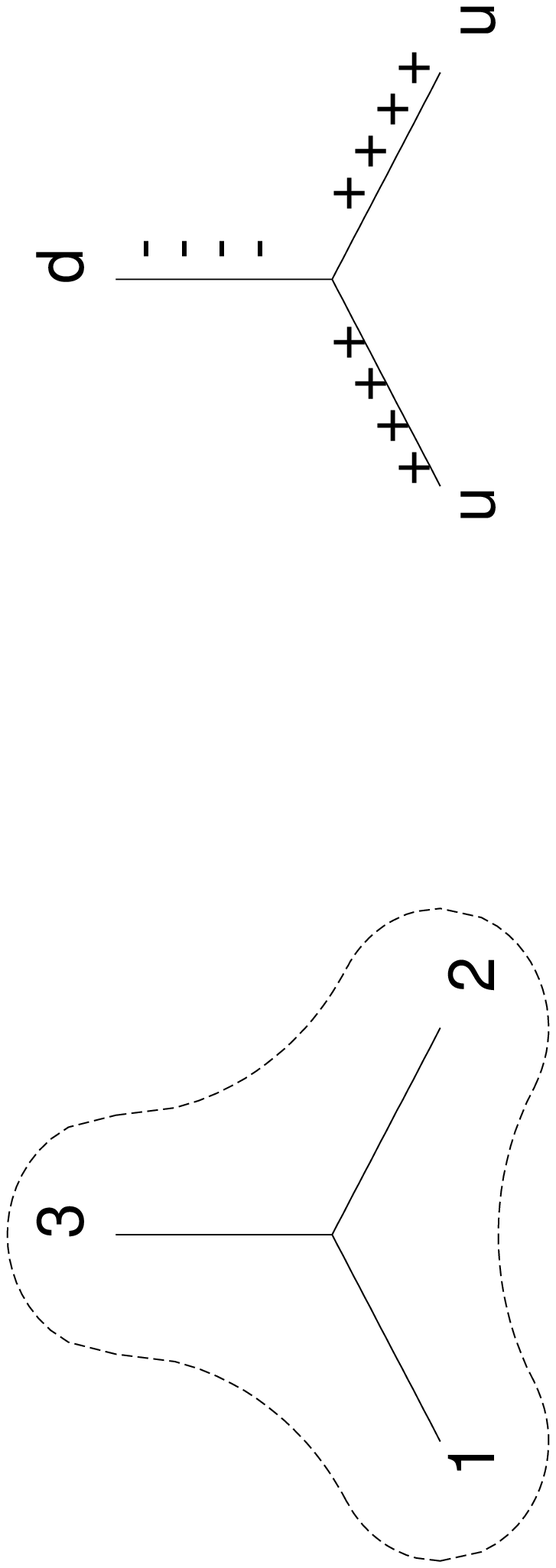}}}}
{\baselineskip=13pt
{\footnotesize{ {\bf Fig. 1.} Collective model of baryons (the charge
distribution of the proton is shown as an example).}}}
\end{tabular}

\noindent\begin{tabular}{p{15cm}}
\epsfxsize 10.0cm
\rotate{\rotate{\rotate{\epsffile{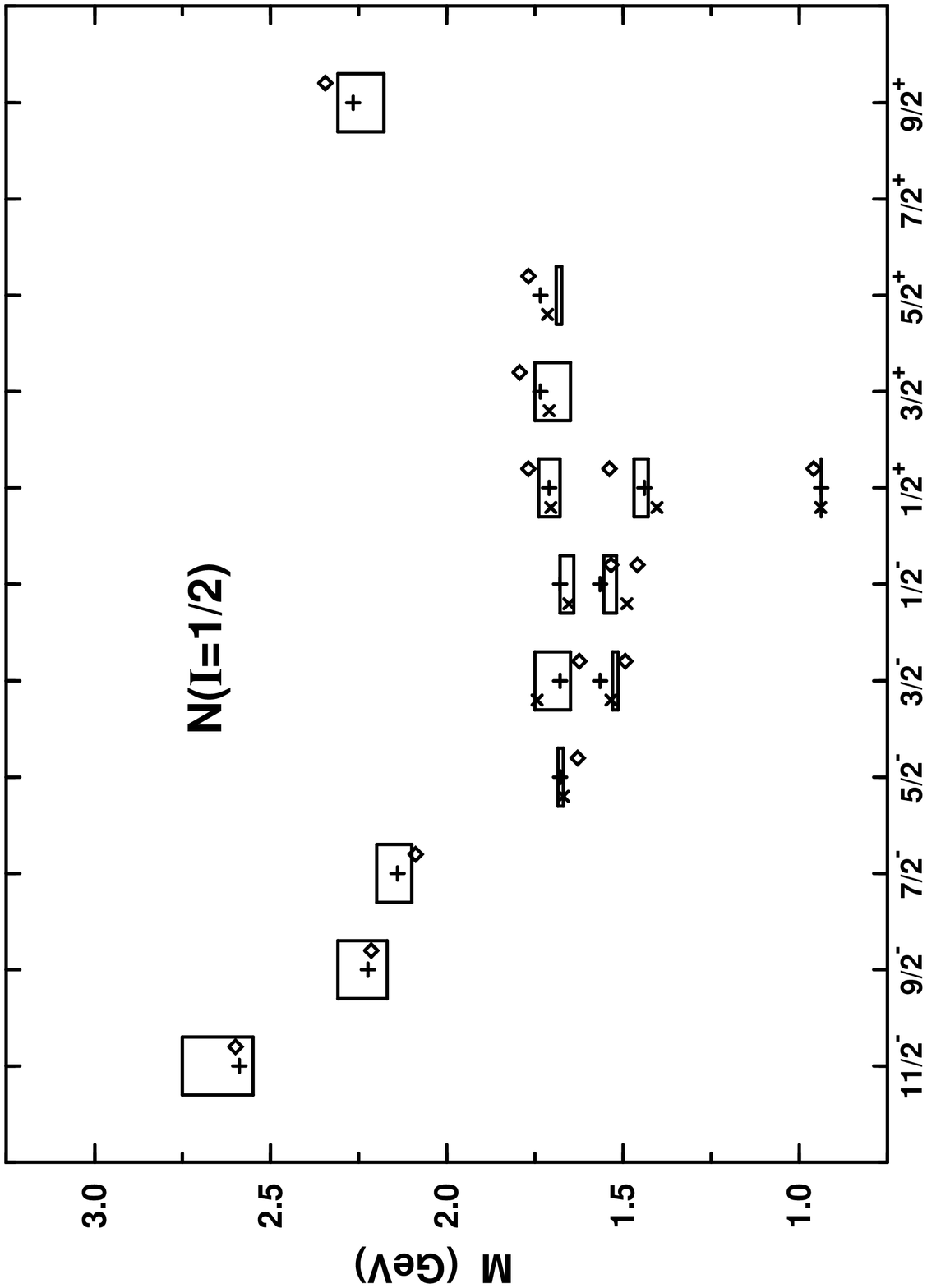}}}}
{\baselineskip=13pt
{\footnotesize{ {\bf Fig. 2.}
Mass spectrum ($M$ {\it vs.} $J^{P}$) for 3* and 4* nucleon resonances.
Collective model\cite{BIL} $(+)$, nonrelativistic quark model\cite{NRQM}
$(\times)$, relativized quark model\cite{RQM} $(\diamond)$.}}}
\vspace{8pt}
\end{tabular}

\noindent
along with a 
comparison to the non-relativistic\cite{NRQM} and relativized
\cite{RQM} quark models. 
As can be seen, the quality of the fits is comparable although the 
underlying dynamics is different. 
This shows that masses alone are not sufficient to distinguish between 
single-particle and collective forms of dynamics and one has to examine
other observables which are more sensitive to the structure of 
wave-functions, such as electromagnetic and strong couplings.

To consider decay processes of baryon resonances in the collective model
we need three ingredients. First, the wave functions of the 
initial and final states. These have the form
\ba
\left| \, ^{2S+1}\mbox{dim}\{SU_f(3)\}_J \,
[\mbox{dim}\{SU_{sf}(6)\},L^P]_{(v_1,v_2);K} \, \right> ~. \label{wf}
\ea
The spin-flavor part has the usual $SU_{sf}(6)$ classification and
determines the permutation symmetry of the state.
The spatial part is 
characterized by the labels: $(v_1,v_2);K,L^P$, where $(v_1,v_2)$
denotes the vibrations (stretching and bending) of the string configuration
in Fig.~1; $K$ denotes the projection of the
rotational angular momentum $L$ on the body-fixed symmetry-axis and 
$P$ the parity.
In this contribution we focus on the nucleon resonances of Fig.~1 
which, with the exception of the $N(1440)P_{11}$ and $N(1710)P_{11}$ 
vibrational states, are associated with rotational excitations of the 
$(v_1,v_2)=(0,0)$ vibrational ground state. 
The second ingredient is the form of the electromagnetic (strong) 
transition operator. It is assumed to involve the absorption or emission 
of a photon (elementary meson) from a single constituent. In such 
circumstances, the couplings discussed below can be expressed 
in terms of the operators
\ba
\hat U &=& \mbox{e}^{ -ik \sqrt{\frac{2}{3}} \lambda_z} ~,
\nonumber\\
\hat T_{m} &=& \frac{i m_{3} k_0}{2} \left(
\sqrt{\frac{2}{3}} \, \lambda_m \,
\mbox{e}^{ -ik \sqrt{\frac{2}{3}} \lambda_z} \, + \,
\mbox{e}^{ -ik \sqrt{\frac{2}{3}} \lambda_z} \,
\sqrt{\frac{2}{3}} \, \lambda_m \right) ~, \label{ut}
\ea
where 
$\lambda_{m}$ ($m=0,\pm$) are Jacobi coordinates and 
$(k_0,\vec{k})$ is the four-momentum of the absorbed quanta.
The form factors of interest are proportional to the matrix elements of
these operators in the wave-functions of Eq.~(\ref{wf}).
The third ingredient which specifies the collective model
is the distribution of the charge and magnetization along the string.
For the present analysis we use the (normalized) distribution 
\ba
g(\beta) &=& \beta^2 \, \mbox{e}^{-\beta/a}/2a^3 ~, \label{gbeta}
\ea
where $\beta$ is a radial coordinate and $a$ is a scale parameter.
The collective form factors are obtained
by folding the matrix elements of $\hat U$ and
$\hat T_{m}$ with this probability distribution
\ba
{\cal F}(k)   &=& \int \mbox{d} \beta \, g(\beta) \,
\langle \psi_f | \hat U   | \psi_i \rangle ~,
\nonumber\\
{\cal G}_m(k) &=& \int \mbox{d} \beta \, g(\beta) \,
\langle \psi_f | \hat T_{m} | \psi_i \rangle  ~.
\label{radint}
\ea
Here $\psi$ denotes the spatial part of the baryon wave function.
In Ref. [3] these form factors are evaluated algebraically and
closed expressions can be derived in the limit of large model space.
The ansatz of Eq.~(\ref{gbeta}) for the probability distribution
is made to obtain the dipole form for the elastic form factor.
The same distribution is used to calculate inelastic form factors 
connecting other final states. 
All collective form factors are found\cite{emff} 
to drop as powers of $k$. 
This property is well-known experimentally and is in contrast with harmonic 
oscillator based quark models in which all form factors fall off 
exponentially.

\section{Electromagnetic Form Factors and Helicity Amplitudes}
\label{sec:emffhelamp}
The elastic electric ($E$) and magnetic ($M$) collective form factors 
are given by
\ba
G^{N}_E &=& 3 \int \mbox{d} \beta \, g(\beta)\,
\langle \, \Psi ;\, M_J=1/2 \,| \,
e_3\, \hat U \, | \, \Psi ;\,
M_J=1/2 \, \rangle ~,
\nonumber\\
G^{N}_M &=& 3 \int \mbox{d} \beta \, g(\beta)\,
\langle \, \Psi ;\, M_J=1/2 \, | \,\mu_3\,
e_3\, \sigma_{3,z}\, \hat U \, | \, \Psi ;\,
M_J=1/2 \, \rangle ~,
\ea
where $\Psi$ denotes the nucleon wave function
$\left| \, ^28^N_{1/2}[56,0^+]_{(0,0);0} \, \right >$ with $N=p\,(n)$ 
for proton (neutron). Further $e_3$,
$\mu_3=eg_3/2m_{3}\,$, $m_3$, $g_3$, $s_3=\sigma_3/2$ are the charge
(in units of $e$: $e_u=2/3$, $e_d=-1/3$), scale magnetic moment, mass,
$g$-factor and spin, respectively, of the third constituent.
Assuming $SU_{sf}(6)$ spin-flavor symmetry ($g_u=g_d$, $\mu_u=\mu_d$),
we obtain the elastic electric form factors
\ba
G_E^p &=& \frac{1}{(1+k^2a^2)^2} \;\;\; ; \;\;\; G_E^n = 0 ~, 
\label{gep}
\ea
and the magnetic form factors are proportional to $G_E^p$ with a ratio
$G_M^n/G_M^p = -2/3$.
Within an effective model with three-constituents,
in order to have a nonvanishing neutron electric form factor,
as experimentally observed, one must break $SU_{sf}(6)$. 
This breaking can be achieved in various ways, {\it e.g.}
by including in the mass operator a hyperfine interaction\cite{iks}, or
by distorting the oblate-top geometry, allowing
for a quark-diquark structure\cite{tzeng}.
Within the model discussed here 
we study the breaking of
the $SU_{sf}(6)$ symmetry by assuming a flavor-dependent distribution
\ba
g_u(\beta) &=& \beta^2 \, \mbox{e}^{-\beta/a_u} /2a_u^3 ~,
\nonumber\\
g_d(\beta) &=& \beta^2 \, \mbox{e}^{-\beta/a_d} /2a_d^3 ~.
\label{gugd}
\ea
In the calculations reported below 
we take $g_u=g_d=1$ and fix $\mu_u$ and $\mu_d$ from the measured 
magnetic moments.
The scale parameters $a_u$ and $a_d$ in the distributions (\ref{gugd})
are determined from a simultaneous fit to the
proton and neutron charge radii, and to the proton and neutron electric 
and magnetic form factors. For the calculations in which the $SU_{sf}(6)$ 
symmetry is satisfied this procedure
yields $a_u=a_d=a=0.232$ fm and
$\mu_u=\mu_d=\mu_p= 2.793$ $\mu_N$ ($=0.126$ GeV$^{-1}$).
When $SU_{sf}(6)$ symmetry is broken
we find $a_u=0.230$ fm, $a_d=0.257$ fm, 
$\mu_u=2.777\mu_{N}$ 
($= 0.126$ GeV$^{-1}$) and $\mu_d=2.915\mu_{N}$ ($=0.133$ GeV$^{-1}$).

Fig.~3 shows the electric and magnetic
form factors of the proton and the neutron.
We see that while the breaking of spin-flavor symmetry can
account for the non-zero value of $G_E^n$ and gives a good description
of the data, it worsens the fit to the proton electric and neutron
magnetic form factors. This implies either that 
the simple mechanism for spin-flavor breaking of Eq.~(\ref{gugd}) does 
not produce the right phenomenology and
other contributions, such as polarization of the neutron into
$p+\pi^{-}$, play an important role in the neutron electric form factor
\cite{cmt}. (A coupling to the meson
cloud through $\rho$, $\omega$ and $\phi$ mesons is indeed expected
\cite{vmd} to contribute in this range of $Q^2$.)
This conclusion ({\it i.e.} worsening the proton form factors)
applies also to the other mechanisms of spin-flavor symmetry
breaking mentioned above, such as that induced by a hyperfine interaction
\cite{iks} which gives $a_u < a_d$ (`moves the up quark to the center and
the down quark to the periphery'). This pattern is a
consequence of the fact that within the framework of constituent models
$G_E^p$, $G_E^n$, $G^p_M$ and $G_M^n$ are intertwined.
\noindent\begin{tabular}{p{5.5cm}p{5.5cm}}
\epsfxsize 5.5cm
\rotate{\rotate{\rotate{\epsffile{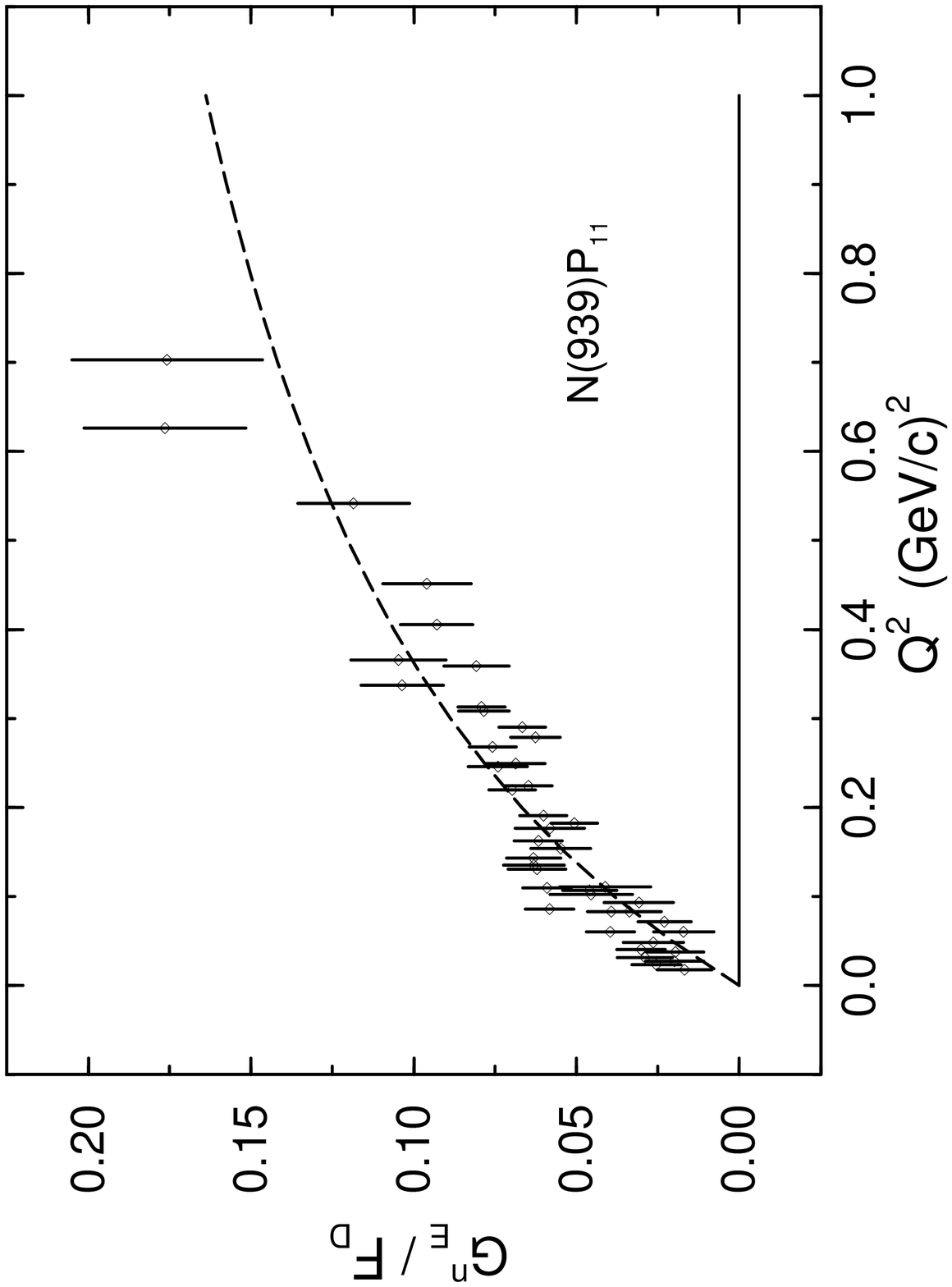}}}}
&
\epsfxsize 5.5cm
\rotate{\rotate{\rotate{\epsffile{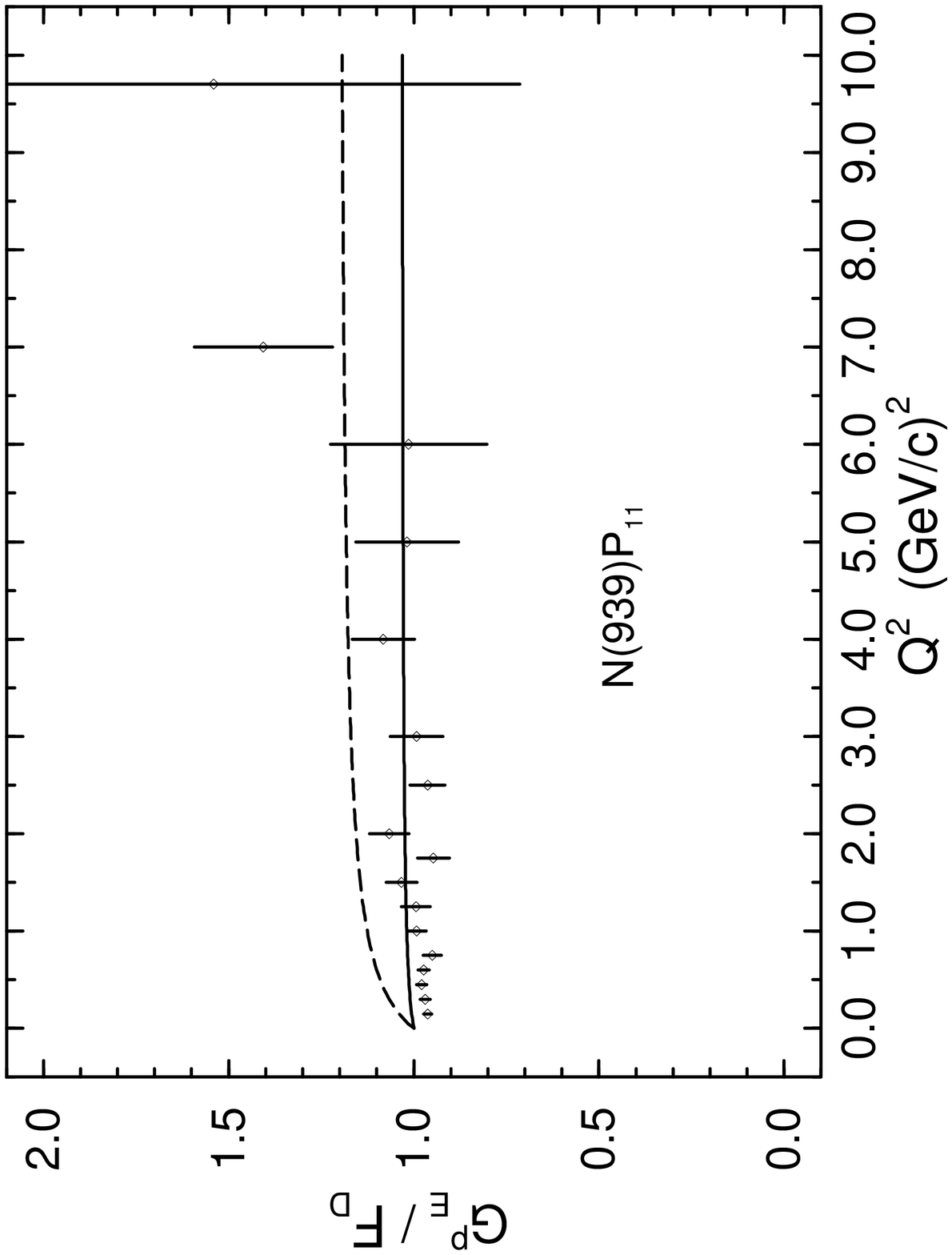}}}}\\
\epsfxsize 5.5cm
\rotate{\rotate{\rotate{\epsffile{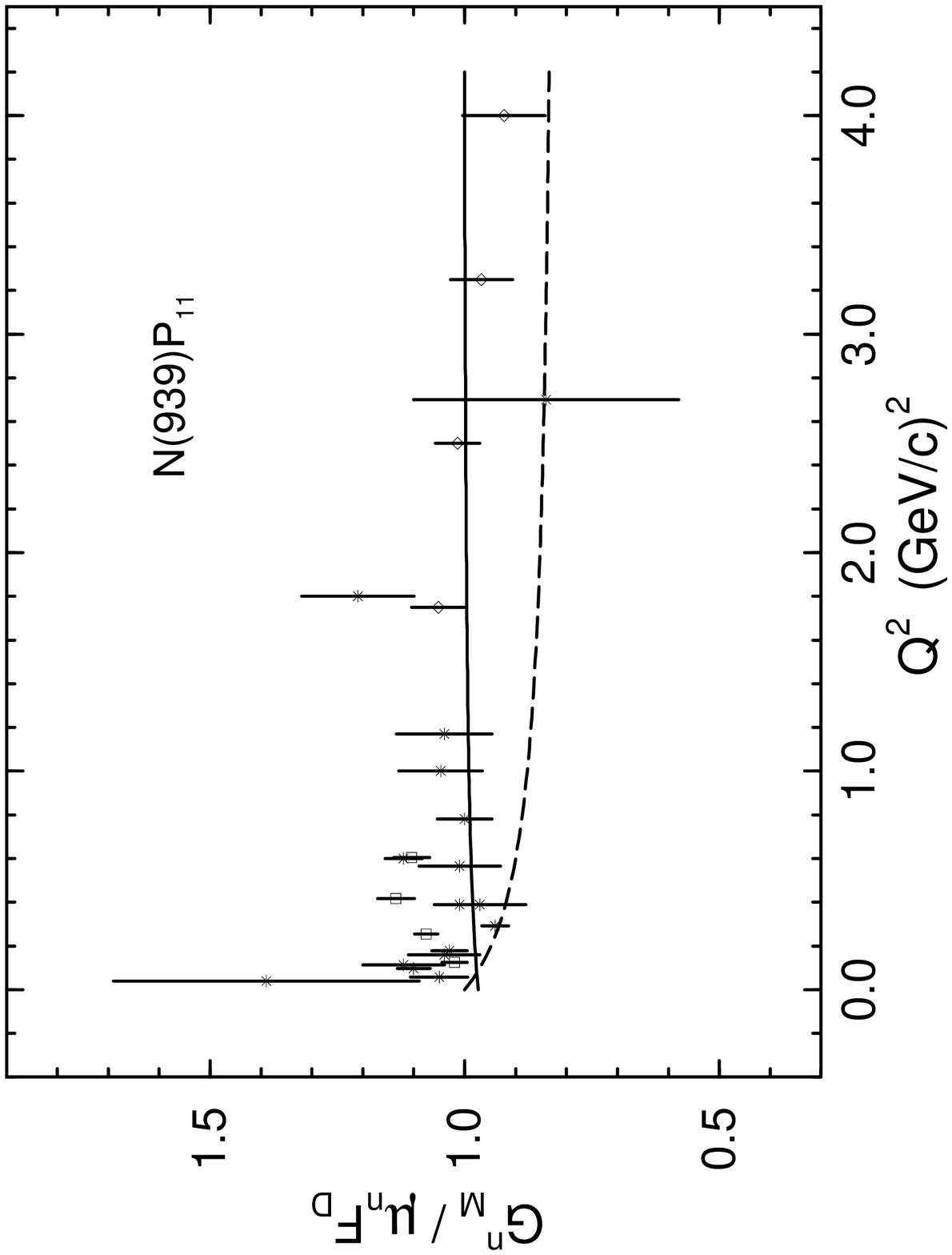}}}}
&
\epsfxsize 5.5cm
\rotate{\rotate{\rotate{\epsffile{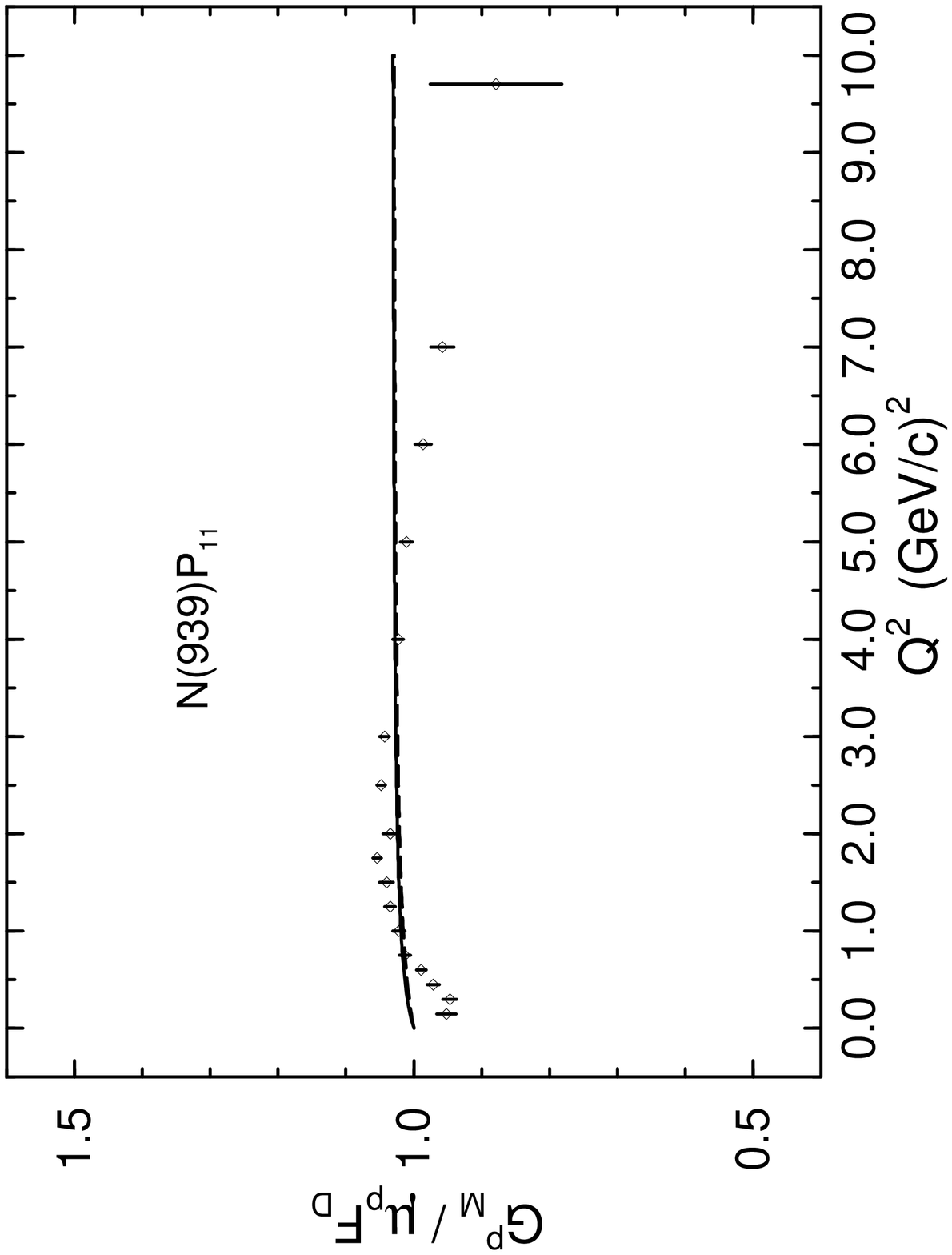}}}}

\end{tabular}
\vspace{3pt}
{\baselineskip=13pt
{\footnotesize{ {\bf Fig. 3.}
Neutron and proton electric ($G_E^n$, $G_E^p$) and magnetic 
($G_M^n/\mu_n$, $G_M^p/\mu_p$) form factors divided by
$F_D=1/(1+Q^2/0.71)^2$. Dashed (solid) lines 
correspond to a calculation with (without) flavor breaking.}}}
\vspace{8pt}

Other (observable) quantities of interest are the helicity amplitudes
in photo- and electroproduction. The transverse helicity amplitudes
between the initial (ground) state of the nucleon and the final
(excited) state of a baryon resonance are 
\ba
A^{N}_{\nu} &=& 6 \sqrt{\frac{\pi}{k_0}} \, \Bigl [ \, k
\langle L,0;S,\nu   | J,\nu \rangle \, {\cal B} -
\langle L,1;S,\nu-1 | J,\nu \rangle \, {\cal A} \, \Bigr ] ~,
\label{helamp}
\ea
where $\nu=1/2$, $3/2$ indicates the helicity.
The orbit- and spin-flip amplitudes (${\cal A}$ and ${\cal B}$,
respectively) are given by
\ba
{\cal B} &=& \int \mbox{d} \beta \, g(\beta)\,
\langle \Psi_f;M_J=\nu | \,\mu_3\,
e_3\, s_{3,+}\, \hat U \, | \Psi_i;M_J^{\prime}=\nu-1 \rangle ~,
\nonumber\\
{\cal A} &=& \int \mbox{d} \beta \, g(\beta)\,
 \langle \Psi_f;M_J=\nu | \, \mu_3\,
e_3\, \hat T_{+}/g_3 \, | \Psi_i;M_J^{\prime}=\nu-1 \rangle ~.
\label{ab}
\ea
These observables correspond to an absorption process 
($\gamma + B^{\prime} \rightarrow B$) of a right-handed 
\noindent\begin{tabular}{p{7.0cm}p{7.0cm}}
\epsfxsize 5.5cm
\rotate{\rotate{\rotate{\epsffile{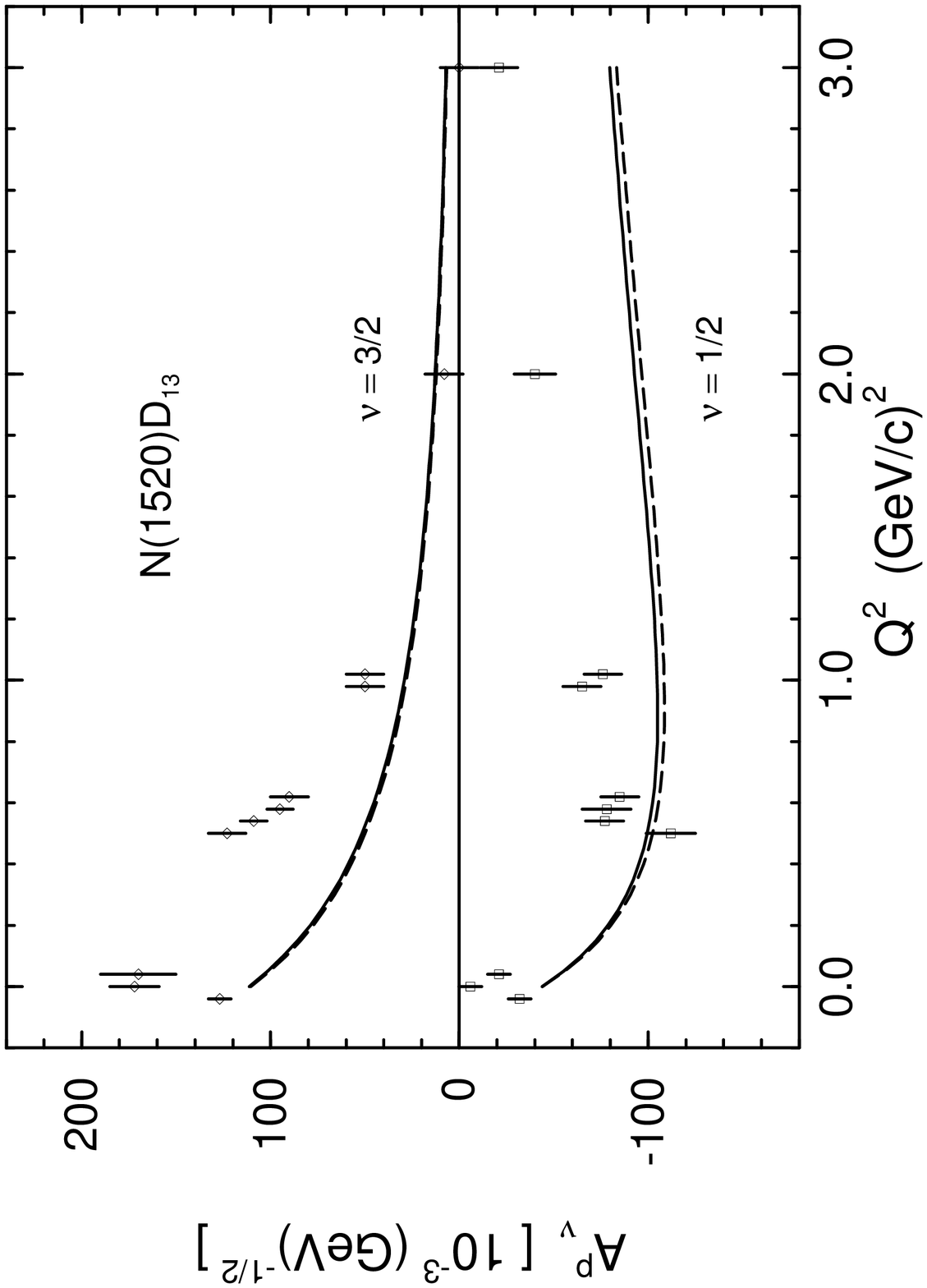}}}}
\vspace{0.5cm}
{\baselineskip=13pt
{\footnotesize{ {\bf Fig. 4.} 
Proton helicity amplitudes for excitation of $N(1520)D_{13}$.
Notation for dashed (solid) lines as in Fig.~3}}}
&
\epsfxsize 5.5cm
\rotate{\rotate{\rotate{\epsffile{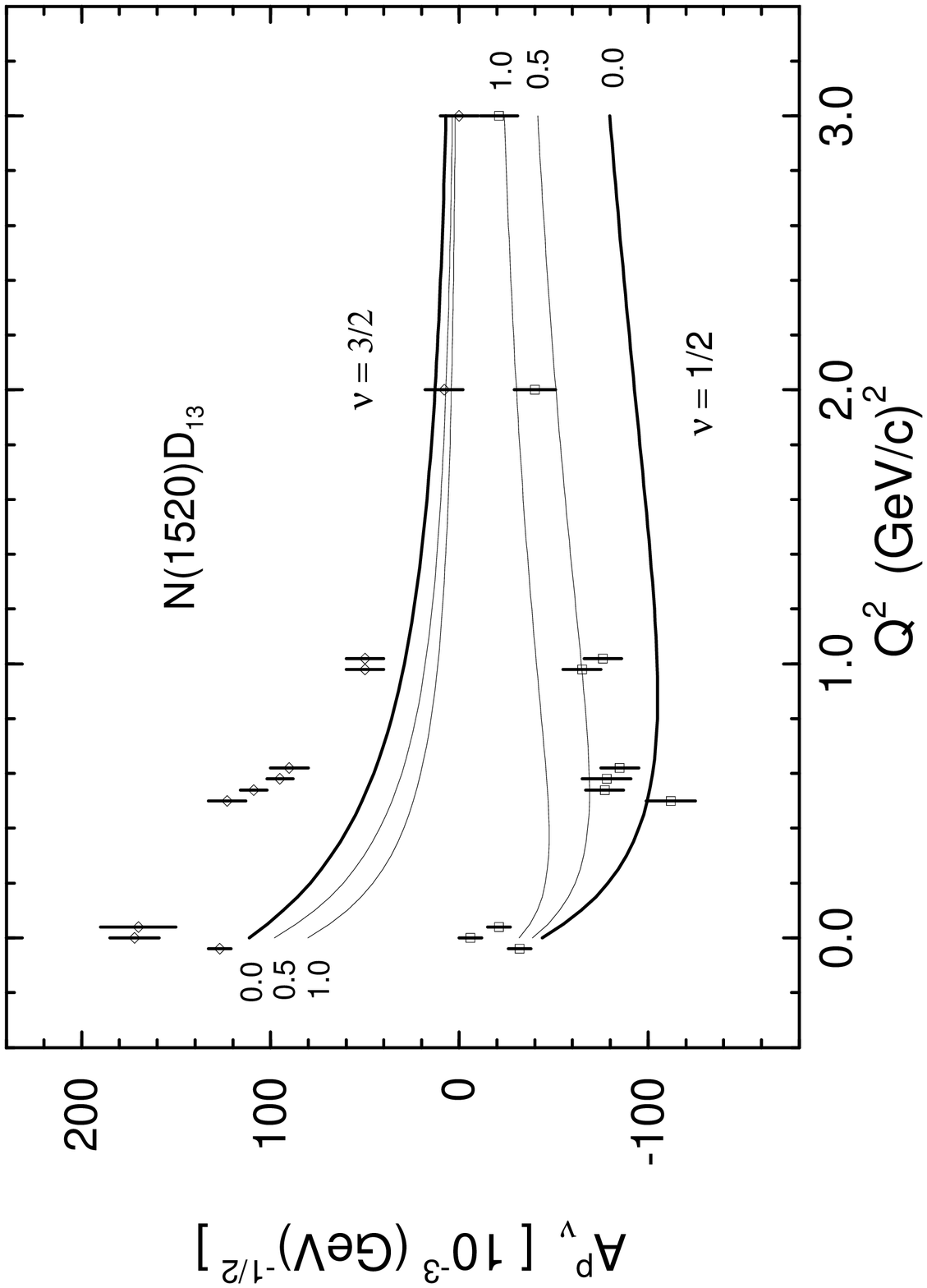}}}}
\vspace{0.5cm}
{\baselineskip=13pt
{\footnotesize{ {\bf Fig. 5.} Effect of hadron swelling for excitation of
$N(1520)D_{13}$. The curves are labelled by the stretching parameter $\xi$
of Eq.~(\ref{stretch}).}}}

\end{tabular}

\noindent
photon with four-momentum ($k_0,\vec{k}=k\hat{z}$). In Eq.~(\ref{ab}) 
$| \Psi_i \rangle$ denotes the (space-spin-flavor) wave function
of the initial nucleon ($B^{\prime}$) with
$\left| \, ^{2}8^N_{1/2}[56,0^+]_{(0,0);0}\,\right >$ and $N=p,n$,
and, similarly,
$| \Psi_f \rangle$ that of the final baryon resonance ($B$).
In general, the ${\cal B}$ and 
${\cal A}$ amplitudes
of Eq.~(\ref{ab}) are proportional to the collective form factors
${\cal F}$ and ${\cal G}_+$ of Eq.~(\ref{radint}), respectively.
When spin-flavor symmetry is broken the helicity
amplitudes are given in terms of flavor-dependent collective form factors
${\cal F}_u(k)$, ${\cal G}_{u,+}(k)$ and ${\cal F}_d(k)$,
${\cal G}_{d,+}(k)$, which depend on the size parameters,
$a_u$ and $a_d$, respectively\cite{emff}. 

The transverse helicity amplitudes $A^{p}_{1/2}$, 
$A^{p}_{3/2}$ in the Breit frame for $N(1520)D_{13}$ 
are shown in Fig.~4 (a factor of $+i$ is suppressed).
As seen, the effect of spin-flavor breaking is rather small.
Only in those cases in which the amplitude with $SU_{sf}(6)$
symmetry is zero, the effect is of some relevance.
Such is the case with proton helicity amplitudes for the
$^{4}8_{J}[70,L^{P}]$ multiplet
({\it e.g.} the $L^{P}=1^{-}$ resonances $N(1675)D_{15}$ and 
$N(1700)D_{13}$) and with neutron $\nu=3/2$ 
amplitudes for the $^{2}8_{J}[56,L^{P}]$ multiplet
({\it e.g.} the $L^{P}=2^{+}$ resonance $N(1680)F_{15}$).

In a string-like model of hadrons one expects\cite{johnsbars} 
on the basis of QCD that strings will elongate (hadrons swell)
as their energy increases. This effect can be
easily included in the present analysis by making the scale parameters
of the strings energy- dependent. We use here the simple ansatz
\ba
a &=& a_0\Bigl ( 1 + \xi\,{W-M\over M}\Bigr ) ~, \label{stretch}
\ea
where $M$ is the nucleon mass and $W$ the resonance mass. This ansatz
introduces a new parameter ($\xi$), the stretchability of the string.
Fig.~5 shows that the effect
of stretching on the helicity amplitudes for 
$N(1520)D_{13}$ is rather large
(especially if one takes the value $\xi\approx 1$ which is suggested
by QCD arguments\cite{johnsbars} and the Regge behavior of nucleon
resonances).
In particular, the data for $N(1520)D_{13}$ (and $N(1680)F_{15}$) 
show a clear indication that the form
factors are dropping faster than expected on the basis of the dipole
form. 

\section{Strong Decay Widths}\label{sec:strongwidth}
In addition to electromagnetic couplings, strong decays of baryons
provide an important, complementary, tool to study their structure.
We consider strong decays of the form
$B \rightarrow B^{\prime} + M$.
The process involves an emission (by one of the constituents in
$B$) of an elementary pseudoscalar meson 
meson ($M=\pi$ or $\eta$) with energy
$k_0=E_M=E_B-E_{B^{\prime}}$ and momentum
$\vec{k}=\vec{P}_M=\vec{P}-\vec{P}^{\prime}=k \hat z$. 
Here $\vec{P}$
and $\vec{P}^{\prime}$ 
are the momenta of the initial ($B$) and final baryon ($B^{\prime}$).
The calculations are performed in the rest frame of $B$ ($P_z=0$).
For decays in which the initial baryon has angular
momentum $\vec{J}=\vec{L}+\vec{S}$ and in which the final baryon is either
the nucleon or the delta,
the (strong) helicity amplitudes in the collective model are
\ba
A_{\nu}(k) 
&=&  \frac{1}{(2\pi)^{3/2} (2k_0)^{1/2}} \left[
\langle L, 0,S,\nu   | J,\nu \rangle \, \zeta_0 Z_0(k) + \frac{1}{2}
\langle L, 1,S,\nu-1 | J,\nu \rangle \, \zeta_+ Z_-(k) \right.
\nonumber\\
&& + \left. \frac{1}{2}
\langle L,-1,S,\nu+1 | J,\nu \rangle \, \zeta_- Z_+(k) \right] ~.
\label{anu}
\ea
The coefficients $\zeta_m$ ($m=0,\pm$) are spin-flavor matrix 
elements\cite{strong} 
and the $Z_m(k)$ are radial matrix elements
\ba
Z_0(k) 
&=& 6 \, [gk - \frac{1}{6}hk] \, {\cal F}(k)^{*}
- 6h \, {\cal G}_z(k)^{*} ~,
\nonumber\\
Z_{\pm}(k) 
&=& -6h \, {\cal G}_{\mp}(k)^{*} ~, \label{radme}
\ea
involving the same collective form factors ${\cal F}(k)$, 
${\cal G}_{\mp}(k)$ discussed in Eq.~(\ref{radint}).
The coefficients $g$ and $h$ denote the strength of two terms
in the transition operator. 
The decay widths for a specific channel are given by 
\ba
\Gamma(B \rightarrow B^{\prime} + M) &=& 2 \pi \rho_f \,
\frac{2}{2J+1} \sum_{\nu>0} | A_{\nu}(k) |^2 ~.
\label{dw}
\ea
where $\rho_f$ is a phase space factor. 
In the algebraic method the widths can be obtained in closed form
which allows us to do a straightforward
and systematic analysis of the experimental data.

\begin{table}[t]
\protect
\tcaption{
$N \pi$ decay widths of 3* and 4* nucleon
resonances in MeV.} 
\label{npi}
\small
\vspace{0.4cm}
\begin{center}
\begin{tabular}{lccccc}
\hline
& & & & & \\
State & Mass & Resonance & $k$(MeV) & $\Gamma$(th) & $\Gamma$(exp) \\
& & & & & \\
\hline
& & & & & \\
$S_{11}$ & $N(1535)$ & $^{2}8_{1/2}[70,1^-]_{(0,0);1}$
& $467$ &  $85$ & $79 \pm 38$ \\
$S_{11}$ & $N(1650)$ & $^{4}8_{1/2}[70,1^-]_{(0,0);1}$
& $547$ &  $35$ & $130 \pm 27$ \\
$P_{13}$ & $N(1720)$ & $^{2}8_{3/2}[56,2^+]_{(0,0);0}$
& $594$ &  $31$ & $22 \pm 11$ \\
$D_{13}$ & $N(1520)$ & $^{2}8_{3/2}[70,1^-]_{(0,0);1}$
& $456$ & $115$ & $67 \pm 9$ \\
$D_{13}$ & $N(1700)$ & $^{4}8_{3/2}[70,1^-]_{(0,0);1}$
& $580$ &   $5$ & $10 \pm 7$ \\
$D_{15}$ & $N(1675)$ & $^{4}8_{5/2}[70,1^-]_{(0,0);1}$
& $564$ &  $31$ & $72 \pm 12$ \\
$F_{15}$ & $N(1680)$ & $^{2}8_{5/2}[56,2^+]_{(0,0);0}$
& $567$ &  $41$ & $84 \pm 9$ \\
$G_{17}$ & $N(2190)$ & $^{2}8_{7/2}[70,3^-]_{(0,0);1}$
& $888$ &  $34$ & $67 \pm 27$ \\
$G_{19}$ & $N(2250)$ & $^{4}8_{9/2}[70,3^-]_{(0,0);1}$
& $923$ &  $7$ & $38 \pm 21$ \\
$H_{19}$ & $N(2220)$ & $^{2}8_{9/2}[56,4^+]_{(0,0);0}$
& $905$ &  $15$ & $65 \pm 28$ \\
$I_{1,11}$ & $N(2600)$ & $^{2}8_{11/2}[70,5^-]_{(0,0);1}$
& $1126$ &  $9$ & $49 \pm 20$ \\
& & & & & \\
\hline
\end{tabular}
\end{center}
\end{table}
We consider here decays with emission of $\pi$ and $\eta$.
The experimental data\cite{PDG} are shown in
Tables~1 and 2, where they are compared
with the results of our calculation. The calculated values depend on the
two parameters $g$ and $h$ in Eq.~(\ref{radme})
and on the scale parameter $a$ of Eq.~(\ref{gbeta}).
In the present analysis we determine these parameters
from a least square fit to the $N \pi$
partial widths (which are relatively well known) with the exclusion
of the $S_{11}$ resonances. For the latter the situation is not clear due
to possible mixing of $N(1535)S_{11}$ and $N(1650)S_{11}$ and
the possible existence of a third $S_{11}$ resonance\cite{LW}. As a
result we find $g=1.164$ GeV$^{-1}$ and $h=-0.094$ GeV$^{-1}$. 
The relative sign is consistent with a previous analysis of the strong
decay of mesons\cite{GIK} and with a derivation from the axial-vector
coupling.
The scale parameter, $a=0.232$ fm, extracted in the present fit
is found to be equal to the value extracted in the calculation of
electromagnetic couplings\cite{emff}.
We keep $g$, $h$ and $a$ equal
for {\em all} resonances and {\em all} decay channels ($N \pi$,
$N \eta$, $\Delta \pi$, $\Delta \eta$). 
In comparing with previous calculations, it should be noted that in the 
calculation in the nonrelativistic quark model\cite{KI}
the decay widths are parametrized by four reduced partial wave amplitudes
instead of the two elementary amplitudes $g$ and $h$. Furthermore,
the momentum dependence of these reduced amplitudes are represented
by constants. The calculation in the relativized quark model\cite{CR}
was done using a pair-creation model for the decay and
involved a different assumption on the phase space factor.
Both the nonrelativistic and relativized quark model calculations
include the effects of mixing induced by the hyperfine interaction,
which in the present calculation are not taken into account.

The calculation of decay widths of 3* and 4* nucleon resonances
into the $N \pi$ channel is found to be in fair agreement with 
experiment (see Table~1). The same holds for the $\Delta \pi$
channel\cite{strong}. These results are to a large
extent a consequence of spin-flavor symmetry. 
There does not seem to be anything unusual in the decays into
$\pi$ and our analysis confirms the results of previous analyses.

\begin{table}[t]
\protect
\tcaption{
$N^{\ast} \rightarrow N \eta$ 
decay widths of (3* and 4*) nucleon resonances in MeV.}
\label{ndeta}
\small
\vspace{0.4cm}
\begin{center}
\begin{tabular}{lcccc}
\hline
& & & & \\
State & Mass & $k$(MeV) & $\Gamma$(th) & $\Gamma$(exp) \\
& & & & \\
\hline
& & & & \\
$S_{11}$ & $N(1535)$ & $182$ &  $0.1$ & $74 \pm 39$ \\
$S_{11}$ & $N(1650)$ & $346$ &  $8$   & $11 \pm  6$ \\
$P_{13}$ & $N(1720)$ & $420$ &  $0.2$ & $$ \\
$D_{13}$ & $N(1520)$ & $150$ &  $0.6$ & $$ \\
$D_{13}$ & $N(1700)$ & $400$ &  $4$   & $$ \\
$D_{15}$ & $N(1675)$ & $374$ & $17$   & $$ \\
$F_{15}$ & $N(1680)$ & $379$ &  $0.5$ & $$ \\
$G_{17}$ & $N(2190)$ & $791$ & $11$   & $$ \\
$G_{19}$ & $N(2250)$ & $831$ &  $9$   & $$ \\
$H_{19}$ & $N(2220)$ & $811$ &  $0.7$ & $$ \\
$I_{1,11}$ & $N(2600)$ & $1054$ & $3$ & $$ \\
& & & & \\
\hline
\end{tabular}
\end{center}
\end{table}

Contrary to the decays into $\pi$,
the decay widths into $\eta$ have some unusual
properties. The calculation gives systematically small values
for these widths (see Table~2). This is due to a combination of 
phase space factors and the structure of the transition operator. 
Due to the difference between the $\pi$ and $\eta$ mass, 
the $\eta$ decay widths
are suppressed relative to the $\pi$ decays. The spin-flavor part is
approximately the same for $N \pi$ and $N \eta$, since $\pi$ and $\eta$
are in the same $SU_f(3)$ multiplet. We emphasize here, that
the transition operator was determined
by fitting the coefficients $g$ and $h$ to the $N \pi$ decays
of the 3* and 4* resonances. Hence the $\eta$ decays are
calculated without introducing any further parameters.

The experimental situation is unclear. The 1992 PDG compilation
gave systematically small widths ($\sim 1$ MeV) for all
resonances except $N(1535)S_{11}$. The 1994 PDG compilation
deleted all $\eta$ widths with the exception of $N(1535)S_{11}$.
This situation persists in the latest PDG compilation\cite{PDG},
where $N(1650)S_{11}$ is now assigned a small but non-zero $\eta$ width.
The results of our analysis suggest that the large $\eta$ width for the
$N(1535)S_{11}$ is not due to a conventional $q^3$ state. One possible
explanation is the presence of another state in the same mass region,
{\it e.g.} a quasi-bound meson-baryon $S$ wave resonance just below
or above threshold, for example $N\eta$, $K\Sigma$ or $K\Lambda$
\cite{Kaiser}. Another possibility is an exotic configuration of four
quarks and one antiquark ($q^{4}\bar{q}$).

\section{Summary and Conclusions}\label{sec:summary}
Within a collective model of the nucleon we have 
analyzed simultaneously elastic
form factors and helicity amplitudes in photo- and electroproduction
and strong decay widths.
The logic of the method is that, by starting from the charge and 
magnetization distribution of the ground state (assuming a dipole form to
the elastic form factor of the nucleon), one can obtain the transition 
form factors to the excited states. In the `collective' model, this 
procedure yields {\it a power dependence of all form factors} (elastic
and inelastic) on $Q^2$.
For electromagnetic couplings we
find that, whereas the breaking
of the spin-flavor symmetry hardly effects the helicity amplitudes,
the stretching of hadrons does have a noticeable influence.                    The disagreement between experimental and theoretical elastic form
factors and helicity amplitudes in the low-$Q^2$ region
$0\leq Q^2\leq 1$ (GeV/c)$^2$ may be due to coupling of the photon
to the meson cloud, {\it i.e.} configurations of the type
$q^3-q\bar{q}$. Since such configurations
have much larger spatial extent than $q^3$, their effects
are expected to drop faster with momentum
transfer $Q^2$ than the constituent form factors.
Within the same collective model we
have performed a calculation of the strong widths for decays into
$\pi$ and $\eta$.
The analysis of experimental data shows that,
while the decays into $\pi$ follow the expected pattern, the decays
into $\eta$ have some unusual features.
Our calculations do not show any indication for a large $\eta$ width,
as is observed for the $N(1535)S_{11}$ resonance. The observed large
$\eta$ width indicates the presence of another configuration, which is
outside the present model space. 
Work on the extension of the formalism to include strange baryons is in
progress.

The results reported in this article are based on work done
in collaboration with F. Iachello (Yale).
The work is supported in part by grant No. 94-00059 from the United 
States-Israel Binational Science Foundation (BSF), Jerusalem, Israel 
(A.L.) and by CONACyT, M\'exico under project 
400340-5-3401E and DGAPA-UNAM under project IN105194 (R.B.).

\end{document}